\documentclass[fleqn,twoside]{article}
\usepackage{espcrc2}

\usepackage{epsfig}
\usepackage[figuresright]{rotating}

\newcommand{\AmS}{{\protect\the\textfont2
  A\kern-.1667em\lower.5ex\hbox{M}\kern-.125emS}}
\newcommand{\preprint}{WUB/10-16}

\title{
\vspace{-3em}{\small\hbox{}\hfill\preprint}\\[2em]
NLO QCD calculations with HELAC-NLO}
       
\author{G.~Bevilacqua\address[Demokritos]{Institute of Nuclear Physics,
               NCSR "Demokritos", GR-15310 Athens, Greece },
             M.~Czakon\address[Aachen]{ Institut f\"ur Theoretische
               Physik E, RWTH Aachen University D-52056 Aachen, Germany},
	    M.~V.~Garzelli\address[Granada]{Departamento de F\'isica Te\'orica y del Cosmos y 
	      CAFPE, E-18071 Granada, Spain}\address[Milano]{INFN Milano, I-20133 Milano, Italy},
	    A.~van~Hameren\address[Cracow]{Institute of Nuclear Physics, Polish Academy 
	      of Sciences, PL-31342 Cracow, Poland},
	    Y.~Malamos\address[Nijmegen]{Department of Theoretical High Energy Physics, Institute
	      for Mathematics, Astrophysics and Particle Physics, Radboud Universiteit Nijmegen, 6525
	       AJ Nijmegen, the Netherlands},
	    C.~G.~Papadopoulos\addressmark[Demokritos],
	    R.~Pittau\addressmark[Granada]
	    and
	    M.~Worek\address{Fachbereich C Physik, Bergische Universit\"at Wuppertal, D-42097 
	    Wuppertal, Germany}
             }

\begin{document}

\begin{abstract}
Achieving a precise description of multi-parton final states is crucial for many analyses at LHC. In this contribution we review the main features of the \texttt{HELAC-NLO} system for NLO QCD calculations.
As a case study, NLO QCD corrections for $t\bar{t} + 2\mbox{ jet}$ production at LHC are illustrated and discussed.
\end{abstract}

\maketitle

\section{Introduction}

With the successful start of collisions at 7 TeV a few months ago, the Large Hadron Collider has put another milestone towards a thorough exploration of the Terascale.
The large energy available will open many multi-particle channels that have being widely investigated~\cite{Binoth:2010ra}. In view of a correct interpretation of the signals of new physics which might be extracted from data, it is of considerable interest to reduce our theoretical uncertainty for the physical processes under study, especially when large QCD backgrounds are involved. In this respect, the need of Next-to-Leading-Order (NLO) corrections for the LHC is unquestionable.

Many efforts have been made in the last few years in developing efficient techniques for one-loop calculations, using both highly-refined traditional methods and the new approaches of OPP and Generalized Unitarity. A significant achievement of this work has been the calculation of complete NLO predictions for several interesting $2 \to 4$ processes at LHC~\cite{KeithEllis:2009bu,Berger:2009ep,Bredenstein:2009aj,Bevilacqua:2009zn,Melnikov:2009wh,Binoth:2009rv,Bredenstein:2010rs,Bevilacqua:2010ve,Berger:2010vm}.

In this contribution we describe the main features of the \texttt{HELAC-NLO} approach to NLO calculations. We provide an overview of the computational framework and briefly summarize some physics results obtained with this system.

\section{Structure of \texttt{HELAC-NLO}}

\subsection{Virtual corrections}

The calculation of the virtual corrections is organized in the framework of the OPP method.

Any one-loop amplitude is characterized by a set of denominators given as products of loop propagators, $\bar{D}_i(\bar{q}) = (\bar{q}+p_i)^2 -m^2_i$, and a polynomial -- in the loop momentum -- function called numerator, $\bar{N}_I(\bar{q})$, such that it can be represented in the form
\begin{equation}\label{OPP}
{\cal A} = \sum_{I \subset \{ 1,\cdots,n\}} \int \frac{\mu^{4-d}\,d^d\bar{q}}{(2\pi)^d}\,\frac{\bar{N}_I(\bar{q})}{\prod_{i \in I} \bar{D}_i(\bar{q})} \,,
\end{equation}
where both the loop momentum $\bar{q}$ and the numerator function are considered to be evaluated in $d$ dimensions.
On the other hand, it is well known that any one-loop amplitude admits a decomposition in terms of a basis of $d$-dimensional scalar integrals. In the limit $d \to 4$, this decomposition can be cast into the form
\begin{eqnarray}\label{OPP}
{\cal A}  &  =  &  \sum_i d_i \mbox{Box}_i +  \sum_i c_i \mbox{Triangle}_i  \nonumber  \\
               &  +  &  \sum_i b_i \mbox{Bubble}_i + \sum_i a_i \mbox{Tadpole}_i + R \,,
\end{eqnarray}
where Box, Triangle, Bubble and Tadpole denote the known scalar loop functions and $R$ is a rational function in the invariants also known as rational term.

The basic idea of OPP is to decompose the numerator of the integrand function, $N(q)$, into a basis of monomials of the propagators appearing in the denominator: the coefficients of the expansion are the $a,b,c,d$ appearing in Eq.\ref{OPP} plus additional, $q$-dependent, coefficients  also known as spurious terms, whose analytical dependence on the loop momentum is known. 
It is therefore possible to evaluate the coefficients of the expansion by computing $N(q)$ a sufficient number of times (equal the number of coefficients to be determined), by appropriate choices of $q$. One possibility is to select those values of the loop momentum that nullify certain sets of propagators -- the so-called unitarity cuts -- and solve the resulting linear system of equations in terms
of the expansion coefficients. If the numerator $N(q)$ is considered in $d=4$ dimensions, it is necessary to decompose the rational terms, $R=R_1+R_2$: we call $R_1$ the 
part of the rational terms that is derived through the reduction starting with a $d=4$ numerator function. $R_2$ is the part of the rational terms that originates from the explicit $(d-4)$ dependence
of the numerator function evaluated in $d$ dimensions. As has been proven in \cite{Ossola:2008xq} $R_2$ can be fully determined by including new interaction-vertices with up to four legs, in
full similarity with the usual UV counter-terms.

On the one hand the method can be fully implemented numerically, resulting to a very fast reduction of arbitrary multi-leg amplitudes. On the other hand, being inherently algebraic, provides exact results (with arbitrary precision) for the coefficients multiplying the one-loop scalar functions as well as for the rational terms.  
The OPP method thus provides a systematic framework to evaluate the cut-constructible part of the amplitude as well as the $R_1$ term, given a numerator function as input (see \cite{Ossola:2006us,Ossola:2008xq} for details).

In the approach we follow \texttt{HELAC-1LOOP}~\cite{vanHameren:2009dr}, based on \texttt{HELAC-PHEGAS}~\cite{Cafarella:2007pc}, automatically generates all the input numerators for the OPP reduction as well as 
all tree-order-like contributions including the new vertices needed for the full determination of the rational part ($R_2$)~\cite{Draggiotis:2009yb,Garzelli:2009is},
as well as UV-renormalization counter-terms. For a detailed description of the algorithm we refer to \cite{vanHameren:2009dr}.

To summarize, our approach for the evaluation of one-loop amplitudes relies upon three basic elements: 1) an implementation of the OPP reduction; 2) a library of scalar loop integrals; 3) an algorithm to automatically evaluate numerator functions and $R_2$ terms. These tasks are solved respectively by the codes \texttt{CutTools}, \texttt{OneLOop} and \texttt{HELAC-1LOOP}~\cite{Ossola:2007ax,vanHameren:2009dr}.

We organize the calculation of virtual corrections using a re-weighting technique and adopt helicity and colour sampling methods in order to optimize the performance of our system~\cite{Bevilacqua:2009zn}.

The treatment of the QCD colour is based on the colour connection representation. 
Given a generic QCD amplitude, ${\cal M}^{i_1\ldots,i_{n_q},a_1,\ldots,a_{n_g}}_{j_1\ldots,j_{n_q}}$, composed by $n_g$ gluons and $n_q$ quarks (antiquarks) with colour indices  $a$ and $i$ ($j$) respectively, it is possible to get a uniform representation for quarks and gluons by contracting  each gluon index $a$ with a corresponding $t^{a}_{i\,j}$:
\begin{equation}\label{colorconn}
{\cal M}^{i_1\ldots,i_{n_q},a_1,\ldots,a_{n_g}}_{j_1\ldots,j_{n_q}}  
\; \to \;
{\cal M}^{i_1\ldots,i_{n_q+n_g}}_{j_1\ldots,j_{n_q+n_g}}
\end{equation}
The resulting amplitude can be now decomposed in terms of colour-stripped partial amplitudes $A_\sigma$:
\begin{equation}\label{colordec}
{\cal M}_{j_1,j_2,\ldots,j_k}^{i_1,i_2,\ldots,i_k}=\sum_\sigma
\delta_{i_{\sigma_{1}},j_1} \delta_{i_{\sigma_{2}},j_2} \ldots
\delta_{i_{\sigma_{k}},j_k} A_\sigma \; ,
\end{equation}
with $k=n_q+n_g$, and $\sigma$ denoting a permutation of the set $\{1,\ldots,k\}$. Each $A_\sigma$ is interpreted as the contribution to the amplitude associated to one possible way of connecting the external particles through colour flow, the connection being established by the indices of the $\delta$'s.
The Feynman rules that allow the calculation of all $A_\sigma$'s have been described in Refs.~\cite{Kanaki:2000ey,Kanaki:2000ms,Cafarella:2007pc}. 

At the level of squared matrix element, the following equivalence holds:
\begin{equation}\label{fullcolorsum}
\sum_{\{i\},\{j\}}|{\cal
M}_{j_1,j_2,\ldots,j_k}^{i_1,i_2,\ldots,i_k}|^2=\sum_{\sigma,\sigma^\prime}A^*_{\sigma}{\cal
C}_{\sigma,\sigma^\prime}A_{\sigma^\prime} \; ,
\end{equation}
where the colour matrix, ${\cal C}_{\sigma,\sigma\prime}$, can be evaluated automatically by counting the number of cycles that are obtained by closing the colour flows of $\sigma,\sigma^\prime$.
Full colour summation is performed using the right-hand side of Eq.~\ref{fullcolorsum}. 

While appealing, the colour-connection representation has a drawback in that the computation becomes less efficient for processes with high parton multiplicities. Indeed, the number of partial amplitudes that have to be evaluated for a given phase-space point increases factorially with the number of the external coloured particles.
A possible way out is given by the colour sampling approach where, following the left hand side of Eq.~\ref{fullcolorsum}, one generates for each phase-space point a random colour assignment and computes the matrix element ${\cal M}_{j_1,j_2,\ldots,j_k}^{i_1,i_2,\ldots,i_k}$ accordingly: due to Eq.\ref{colordec}, only the subset of the $A_\sigma$'s which have a colour flow compatible with the given assignment contribute. In fact,  as shown in Table \ref{table:1}, the number of subamplitudes that have to be computed in this way is sensibly reduced, on the average, for increasing complexity of the colour structure.
\begin{table*}[htb]
\caption{Comparison between the number of subamplitudes to be evaluated using full colour sum ($n_{conn}$) and using the Monte Carlo approach described in the text (average number, $\langle n_{conn} \rangle_{MC}$), for three reference processes.}
\newcommand{\cc}[1]{\multicolumn{1}{c}{#1}}
\renewcommand{\tabcolsep}{2pc}
\renewcommand{\arraystretch}{1.2}
\begin{tabular}{@{}llll}
\hline
Process   &  \cc{$n_{conn}$} & \cc{$\langle n_{conn} \rangle_{MC}$} & \cc{Ratio} \\
\hline
$gg \to b \bar{b} W^+ W^- $ & 6 & 1.74 & 3.5 \\
$gg \to t \bar{t} b \bar{b} $ & 24 & 3.04 & 7.89 \\
$gg \to t \bar{t} gg $ & 120 & 6.27 & 19.1 \\
\hline
\end{tabular}\\[2pt]
\label{table:1}
\end{table*}

\subsection{Real corrections}
The singularities arising from soft and collinear divergences in the initial and final states are treated within the framework of Catani-Seymour dipole subtraction~\cite{Catani:1996vz}, in its formulation for massive quarks~\cite{Catani:2002hc} and arbitrary polarizations~\cite{Czakon:2009ss}. After combining virtual and real corrections, singularities connected to collinear configurations in the final state as well as soft divergencies in the initial and final state automatically cancel for infrared-safe observables when the jet algorithm is applied. Singularities connected to collinear splittings in the initial state are removed via PDF factorization.

Calculations are performed with the help of the code \texttt{HELAC-DIPOLES}~\cite{Czakon:2009ss} which implements automatic construction of all relevant subprocesses and dipoles as well as  phase-space integration of subtracted real radiation and of the corresponding integrated dipoles. The phase-space integration is performed using the approaches of the multichannel generator 
\texttt{PHEGAS}~\cite{Papadopoulos:2000tt} and of the recently published adaptive-multichannel generator \texttt{KALEU}~\cite{vanHameren:2010gg}.

A phase-space restriction is applied on the contribution of dipoles, as originally proposed in \cite{Nagy:1998bb,Nagy:2003tz}, considering two values of the unphysical cutoff: $\alpha_{max}=1$ and $\alpha_{max}=0.01$. The $\alpha_{max}$ independence is explicitly checked in all our results as a consistency check of the calculation.

\section{Predictions for the LHC: $t\bar{t} + 2\mbox{ jets}$}

The \texttt{HELAC-NLO} framework has been applied in the computation of NLO QCD corrections to several processes relevant for top quark and Higgs phenomenology~\cite{Binoth:2010ra,Bevilacqua:2009zn,Bevilacqua:2010ve}.
We report here, as an example, a brief summary of our results about inclusive production of $t\bar{t}$ in association with two jets, which represents an important background for Higgs searches in the channel $t\bar{t}H (H \to b\bar{b})$~\cite{ATLAS}.

To be specific, we consider proton-proton collisions at the center of mass energy $\sqrt{s}=14$ TeV.
Top quarks are considered on shell without any kinematical restriction applied. For the top quark mass we set $m_t = 172.6$ GeV, whereas all the oher partons are considered massless.
We consistently use the CTEQ6 set of parton distribution functions throughout the calculation, \textit{i.e.} we take the CTEQ6L1 set  with a one-loop running $\alpha_S$ at LO and CTEQ6M pdf's  with a two-loop running $\alpha_S$ for the evaluation of the NLO corrections.

Our definition of jets is based on the $k_T$ algorithm~\cite{Catani:1992zp,Catani:1993hr} using $\Delta R=0.8$, where $\Delta R$ denotes the separation in the azimuthal angle - pseudorapidity plane. Jets are obtained by recombining at most two partons and the recombination takes place only when both partons satisfy $\vert y_i \vert < 5$ (approximate detector bounds). Also $b$ quarks enter our jet definition, so the final state $t\bar{t}b\bar{b}$ is included in the analysis.
We further impose a minimum $\Delta R = 1.0$ separation between jets and require the latter to 
lie within the region defined by $\vert y_{jet} \vert < 4.5$. Finally, we require hard jets to have a minimum transverse momentum of $50$ GeV.

The numerical results of this study are summarized in Table \ref{table:2}, where we show a comparison of the total LO and NLO cross sections at the central scale $\mu_R = \mu_F = m_t$. As a cross check, we report the NLO results obtained for two different values of the dipoles cutoff parameter, $\alpha_{max}=1$ and $\alpha_{max}=0.01$, showing that they are in mutual agreement within the integration error.

\begin{figure}[htb]
\includegraphics[width=68mm,height=55mm]{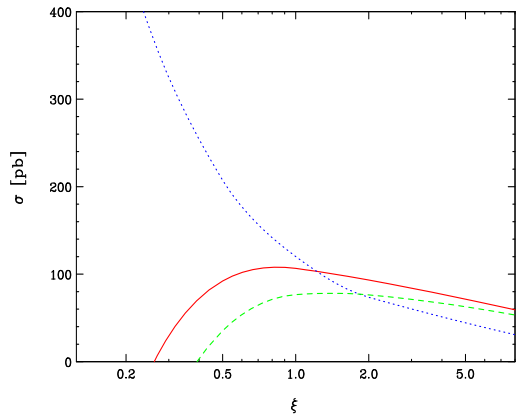}

\caption{Scale dependence of the total cross section for $pp \to t\bar{t} + 2 \mbox{ jets}+X$ at the LHC with $\mu_R = \mu_F = \xi \dot \mu_0$, where $\mu_0 = m_t$. The blue dotted curve corresponds to the LO, the red solid to the NLO result whereas the green dashed to the NLO results with a jet veto of 50 GeV.}
\label{fig:scale_dependence}
\end{figure}

The scale dependence of the corrections is illustrated in Figure \ref{fig:scale_dependence}.
In the absence of any jet veto, QCD corrections determine a negative shift of approximately 11\% compared to the LO results. A dramatic reduction of the scale uncertainty is observed when going from LO to NLO: varying the scale up and down by a factor 2 changes the LO cross section by +72\% and -39\% respectively, while in the NLO case one observes a variation of -13\% and -12\%.
Applying a jet veto of 50 GeV results in an increasing negative shift operated by QCD corrections and leads to a K-factor of 0.64.

\begin{figure}[htb]
\includegraphics[width=68mm,height=55mm]{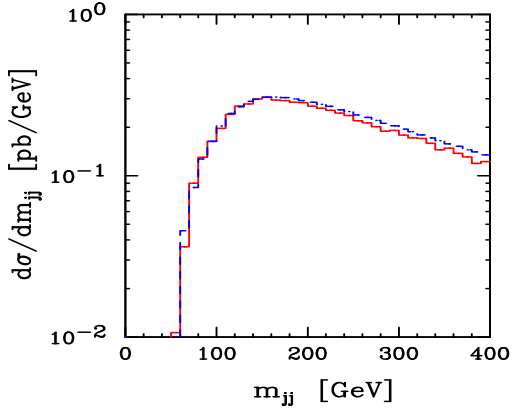}
\caption{Distribution of the di-jet invariant mass for $pp \to t\bar{t} + 2 \mbox{ jets}+X$ at the LHC. The red solid line refers to the NLO results while the blue dotted one to the LO one.}
\label{fig:distributions1}
\end{figure}

Let us now analyse the changes in shape of distributions determined by QCD corrections. 
A few differential cross sections for simple observables are considered. The most important one for us is the di-jet invariant mass since it is the observable entering Higgs boson studies. This is illustrated in Figure \ref{fig:distributions1}, where we focus the attention in the range up to 400 GeV due to our phenomenological motivations. We notice that the distribution shows tiny corrections up to at least 200 GeV. 

\begin{figure}[htb]
\includegraphics[width=68mm,height=55mm]{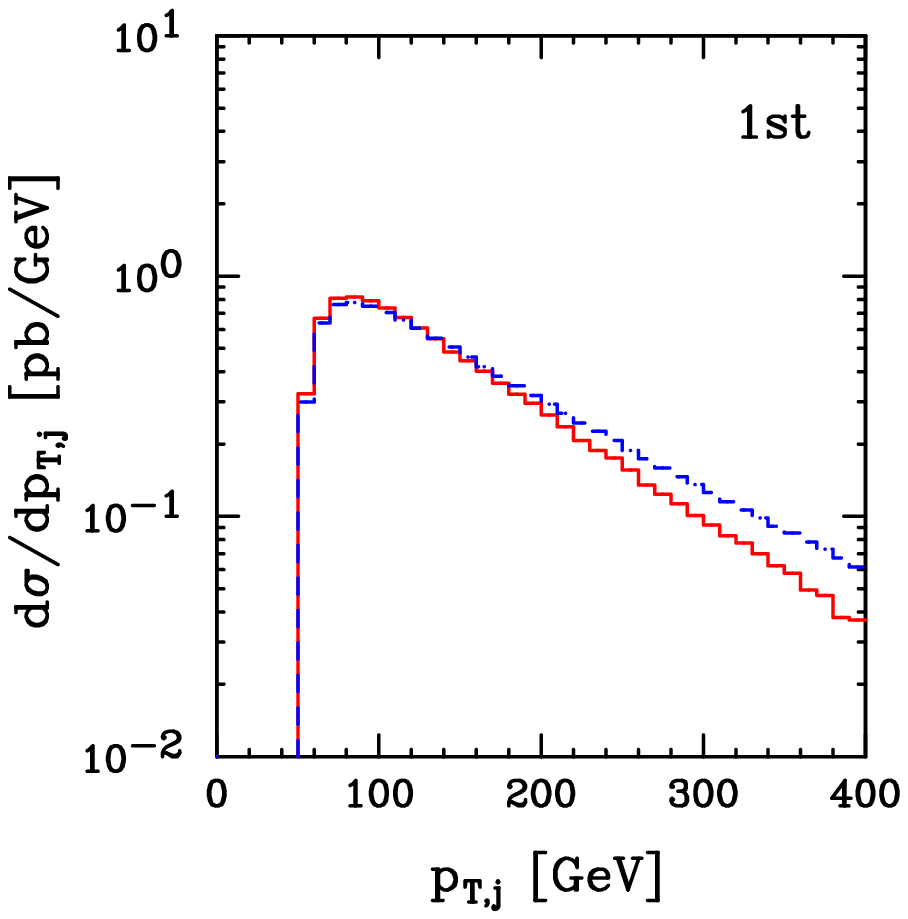}
\caption{Distributions of the transverse momentum for the first hardest jet for $pp \to t\bar{t} + 2 \mbox{ jets}+X$ at the LHC. The red solid line refers to the NLO results while the blue dotted one to the LO one.}
\label{fig:distributions2}
\end{figure}

\begin{figure}[htb]
\includegraphics[width=68mm,height=55mm]{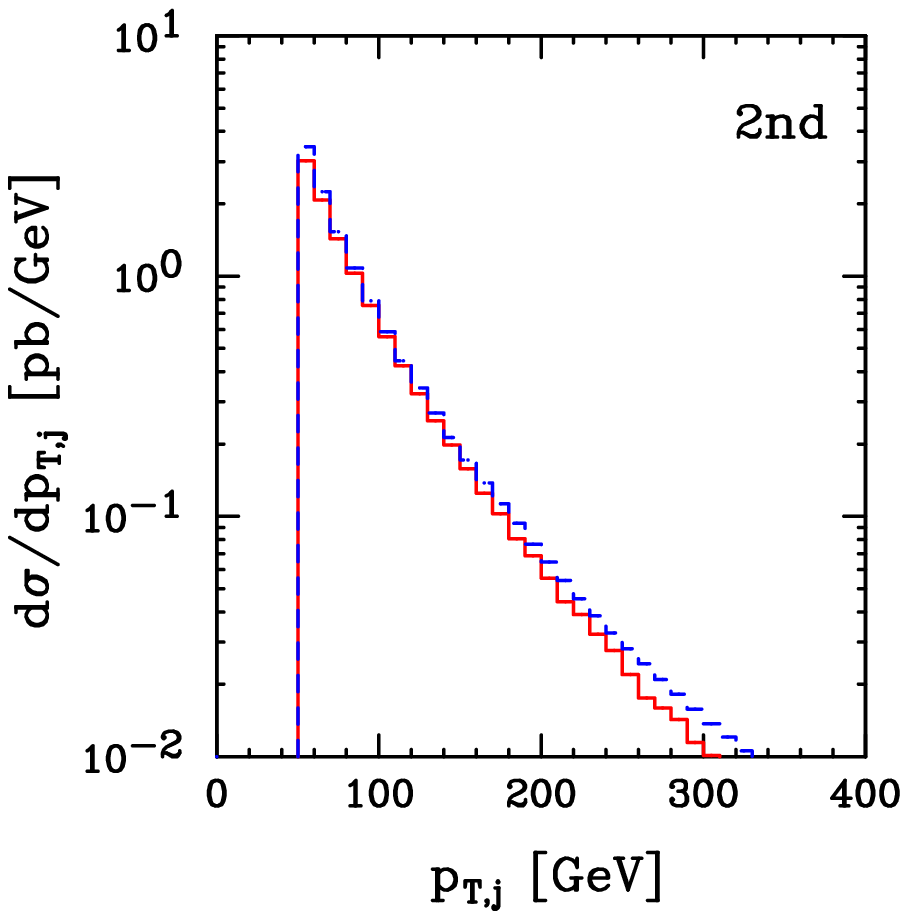}
\caption{Distributions of the transverse momentum for the second hardest jet for $pp \to t\bar{t} + 2 \mbox{ jets}+X$ at the LHC. The red solid line refers to the NLO results while the blue dotted one to the LO one.}
\label{fig:distributions3}
\end{figure}

Much larger effects can be found in the other observables analysed, namely the transverse momentum of the first and of the second hardest jet. Figures \ref{fig:distributions2} and \ref{fig:distributions3} show visible differences in shape in the region of high $p_T$, especially in the case of the first hardest jet. We conclude by pointing out that an accurate description of the tails of these $p_T$ distributions cannot prescind from the inclusion of higher-order corrections. On the other hand, the behaviour at low $p_T$'s is altered by soft collinear emissions whose effects can be simulated by parton showers.
\begin{table*}[htb]
\caption{Cross section at LO and NLO for $t\bar{t} + 2 \mbox{ jets}$ production at LHC ($\sqrt{s}=14$ TeV).The two NLO results reported refer to different values of the dipole phase-space cutoff $\alpha_{max}$. The scale choice is $\mu_R=\mu_F=m_t$.
}
\label{table:2}
\newcommand{\cc}[1]{\multicolumn{1}{c}{#1}}
\renewcommand{\tabcolsep}{2pc}
\renewcommand{\arraystretch}{1.4}
\begin{tabular}{@{}llll}
\hline
$pp \to t\bar{t}+2\mbox{ jets} + X$  &  \cc{$\sigma^{\mbox{\tiny{LO}}}$ (pb)} & \cc{$\sigma^{\mbox{\tiny{NLO}}}_{\alpha_{max}=1}$ (pb)} & \cc{$\sigma^{\mbox{\tiny{NLO}}}_{\alpha_{max}=0.01}$ (pb)} \\
\hline
no jet veto  & $120.17 \pm 0.08$ & $106.95 \pm 0.17$ & $106.56 \pm 0.31$  \\
with jet veto ($p_{T,X} < 50$ GeV)   & $120.17 \pm 0.08$  &  $76.58 \pm  0.17$  &  $76.57 \pm  0.37$ \\
\hline
\end{tabular}\\[2pt]
\end{table*}
\section{Outlook}

It is amazing how much progress has been realized over the last few years in the field of multi-leg NLO calculations. \texttt{HELAC-NLO} is a complete framework for NLO calculations, including virtual and real corrections. In its present form can be used for NLO QCD calculations for arbitrary processes 
including up to one-particle-irreducible 8-point one-loop functions.    
It has been successfully used for a  number of processes of current interest, such as
$pp\to t\bar{t} b\bar{b}$ including the Higgs signal contributions and $pp\to t\bar{t} +2$ jets.  In the immediate future, we plan a complete analysis of $t\bar{t}+$jets production including
the full description of top decays and non-resonant contributions, i.e. $pp\to l\bar{\nu} \bar{l}^\prime\nu^\prime b\bar{b}$. Plans of further development include also the incorporation of all $R_2$ and UV counter-terms
for  full electroweak corrections.

%\section{Acknowledgement}

{\scriptsize \textit{Acknowledgement: This work was funded in part by the RTN European Programme 
MRTN-CT-2006-035505 HEPTOOLS - Tools and Precision Calculations 
for Physics Discoveries at Colliders. M.C. was  supported by the 
Heisenberg Programme of the Deutsche Forschungsgemeinschaft. M.W. was 
supported by the Initiative and Networking Fund of the Helmholtz 
Association, contract HA-101 ("Physics at the Terascale"). 
M.V.G. and R.P. thank the financial support of the MEC project FPA2008-02984.
M.V.G. was additionally supported by the INFN.}
}

\end{document}